\documentclass[aps,twocolumn,prl,tightenlines,floatfix,superscriptaddress,longbibliography]{revtex4-2}

\usepackage[breaklinks=true]{hyperref}
\usepackage{graphicx}
\usepackage[english]{babel}
\usepackage{amsmath}
\usepackage{amssymb}
\usepackage{times}
\usepackage{braket}

\begin{document}

\title{Quantum Geometry, Anomalous Scaling, and Strong Pseudogap Superfluidity in a Flat-Band Lieb Lattice}

\author{Lin Sun}
\email{These authors contributed equally to this work.}
\affiliation{Hefei National Laboratory, University of
 Science and Technology of China, Hefei 230088, China}
\affiliation{Shanghai Research Center for Quantum Science and CAS Center for Excellence in Quantum Information and Quantum Physics, 
University of Science and Technology of China, Shanghai 201315, China}

\author{Hao Deng}
\email{These authors contributed equally to this work.}
\affiliation{Hefei National Research Center for Physical Sciences at the Microscale and School of Physical Sciences, University 
   of Science and Technology of China, Hefei, Anhui 230026, China}
\affiliation{Shanghai Research Center for Quantum Science and CAS Center for Excellence in Quantum Information and Quantum Physics, 
 University of Science and Technology of China, Shanghai 201315, China}
\affiliation{Hefei National Laboratory, University of
  Science and Technology of China, Hefei 230088, China}
\affiliation{Key Laboratory of Artificial Structures and Quantum Control, School of Physics and Astronomy, Shanghai Jiao Tong University, Shanghai 200240, China}

\author{Yuxuan Wu}
\affiliation{Hefei National Research Center for Physical Sciences at the Microscale and School of Physical Sciences, University 
   of Science and Technology of China, Hefei, Anhui 230026, China}
\affiliation{Shanghai Research Center for Quantum Science and CAS Center for Excellence in Quantum Information and Quantum Physics, 
 University of Science and Technology of China, Shanghai 201315, China}
\affiliation{Hefei National Laboratory, University of
  Science and Technology of China, Hefei 230088, China}
\author{Chuping Li}
\affiliation{Hefei National Research Center for Physical Sciences at the Microscale and School of Physical Sciences, University 
   of Science and Technology of China, Hefei, Anhui 230026, China}
\affiliation{Shanghai Research Center for Quantum Science and CAS Center for Excellence in Quantum Information and Quantum Physics, 
 University of Science and Technology of China, Shanghai 201315, China}
\affiliation{Hefei National Laboratory, University of
  Science and Technology of China, Hefei 230088, China}
\author{Junru Wu}
\affiliation{Hefei National Research Center for Physical Sciences at the Microscale and School of Physical Sciences, University 
   of Science and Technology of China, Hefei, Anhui 230026, China}
\affiliation{Shanghai Research Center for Quantum Science and CAS Center for Excellence in Quantum Information and Quantum Physics, 
 University of Science and Technology of China, Shanghai 201315, China}
\affiliation{Hefei National Laboratory, University of
  Science and Technology of China, Hefei 230088, China}
\author{Kaichao Zhang}
\affiliation{Hefei National Research Center for Physical Sciences at the Microscale and School of Physical Sciences, University 
   of Science and Technology of China, Hefei, Anhui 230026, China}
\affiliation{Shanghai Research Center for Quantum Science and CAS Center for Excellence in Quantum Information and Quantum Physics, 
 University of Science and Technology of China, Shanghai 201315, China}
\affiliation{Hefei National Laboratory, University of
  Science and Technology of China, Hefei 230088, China}
\author{Pengyi Chen}
\affiliation{Hefei National Research Center for Physical Sciences at the Microscale and School of Physical Sciences, University 
   of Science and Technology of China, Hefei, Anhui 230026, China}
\affiliation{Shanghai Research Center for Quantum Science and CAS Center for Excellence in Quantum Information and Quantum Physics, 
 University of Science and Technology of China, Shanghai 201315, China}
\affiliation{Hefei National Laboratory, University of
  Science and Technology of China, Hefei 230088, China}
\author{Dingli Yuan}
\affiliation{Hefei National Research Center for Physical Sciences at the Microscale and School of Physical Sciences, University 
   of Science and Technology of China, Hefei, Anhui 230026, China}
\affiliation{Shanghai Research Center for Quantum Science and CAS Center for Excellence in Quantum Information and Quantum Physics, 
 University of Science and Technology of China, Shanghai 201315, China}
\affiliation{Hefei National Laboratory, University of
  Science and Technology of China, Hefei 230088, China}

\author{Qijin Chen}
\email[Corresponding author: ]{qjc@ustc.edu.cn}
\affiliation{Hefei National Research Center for Physical Sciences at the Microscale and School of Physical Sciences, University 
   of Science and Technology of China, Hefei, Anhui 230026, China}
\affiliation{Shanghai Research Center for Quantum Science and CAS Center for Excellence in Quantum Information and Quantum Physics, 
 University of Science and Technology of China, Shanghai 201315, China}
\affiliation{Hefei National Laboratory, University of
  Science and Technology of China, Hefei 230088, China}

\date{\today}

\begin{abstract}
  Flat-band systems such as magic-angle twisted bilayer graphene host
  strong-correlation superconductivity at vanishingly weak coupling,
  yet how quantum geometry and pairing fluctuations conspire to drive
  this phenomenon remains an open question.  We investigate
  finite-temperature superfluidity in a quasi-two-dimensional Lieb
  lattice using a pairing fluctuation theory with a band-uniform
  attractive interaction $g<0$ that isolates the intrinsic quantum
  geometric contributions.  Quantum geometry significantly amplifies
  superfluidity; the geometric pair hopping integral surpasses its
  conventional counterpart, and the geometric superfluid density
  becomes the dominant in-plane transport component.  When the Fermi
  level enters the flat band, the BCS paradigm breaks down
  entirely; the pairing gap and $T_\text{c}$ shift from exponential to
  anomalous power-law scaling $\Delta, T_\text{c} \propto |g|^\nu$ ($\nu>1$),
  and the superfluid density inherits an unconventional power-law
  temperature dependence at low temperatures.  In the 2D limit
  ($t_z=0$), the pseudogap at $T_\text{c}$ nearly saturates the
  zero-temperature gap even at $|g|/t=0.001$, placing the system in a
  strong-pseudogap regime that would otherwise require unitary or
  BEC-scale interactions.  These findings establish a microscopic
  mechanism for flat-band enhanced superfluidity and offer testable
  predictions for ultracold atom experiments.
\end{abstract}

\maketitle

Ultracold Fermi gases in optical lattices constitute a versatile
platform for exploring strongly correlated quantum many-body physics,
offering unprecedented tunability over key system parameters such as
interaction strength, lattice depth, temperature, dimensionality, and
population imbalance
~\cite{Chen2005PR,Bloch2008RMP,Hart2015N,Demarco1999S,Bartenstein2004PRL,Kinast2005S,zwierlein2006S,partridge2006S}.
This precise control not only facilitates the exploration of the
BCS-BEC crossover in fermionic superfluids~\cite{Chin2010RMP} but also
enables the simulation of fundamental condensed matter paradigms, most
notably the Hubbard model~\cite{Jaksch1998PRL,Bloch2008N}.  Such
capabilities provide critical insights into strongly correlated
phenomena, including the widespread pseudogap behavior and the elusive
mechanisms governing high-temperature
superconductivity~\cite{Chen1998PRL,Timusk1999RPP,Li2024N}.

In the search for exotic quantum states, the Lieb
lattice~\cite{Lieb1989PRL} has attracted significant attention as a
unique bipartite geometry featuring a dispersionless flat band
characterized by a singular density of states (DOS) and a zero Chern
number~\cite{DENG2024AoP,Chen2014JPAMT}.  Its unit cell contains three
sites (one at the corner and two at the midpoints of adjacent edges),
resulting in a band structure where a central flat band contacts two
dispersive bands at Dirac points~\cite{Huhtinen2018PRB}.  This
distinctive topology quenches the kinetic energy, thereby amplifying
correlation effects and hence the possibility of high-temperature
superconductivity.  While experimental realizations in ultracold
atoms~\cite{Taie2015SA,Schafer2020NRP,Cui2020NC} have enabled studies
of magnetism and topological quantum spin Hall
states~\cite{Noda2009PRA,Noda2015PRA,Nie2017PRA,Zhu2017PRB,Sadeghi2022SR,Beugeling2012PRB,Pires2022JMMM},
the role of the flat band's quantum geometry in superfluidity remains
a frontier of research.  Recent studies in various flat-band systems,
including checkerboard lattices, magic-angle twisted graphene
superlattices, kagome lattices, and honeycomb
lattices~\cite{Sun2011PRL,Cao2018N,Tang2011PRL,Neupert2011PRL},
suggest that quantum geometric effects may enhance the superconducting
transition temperature and lead to quantum Hall states with nonzero
Chern
numbers~\cite{Kopnin2011PRB,Kauppila2016PRB,Wang2011PRL,Beugeling2012PRB}.
In the context of the Lieb lattice, such geometric enhancement is
highlighted by recent ground-state calculations, showing that the flat
band induces anomalous interaction-dependent power laws in both the
pairing gap and superfluid density~\cite{DENG2024AoP}. It is also
shown that such flat-band induced superfluidity is robust against
disorders~\cite{Bouzerar2025}. Yet, understanding how the interplay
between the flat band's quantum geometry and pairing fluctuations
governs finite-temperature superfluidity remains a fundamental open
question.

In this Letter, we investigate the superfluidity and pairing phenomena
of ultracold Fermi gases in a quasi-two-dimensional Lieb lattice.  To
disentangle the intrinsic quantum geometric contributions from
orbital-dependent interaction effects, we employ an effective
Hamiltonian with a band-uniform attractive interaction using a pairing
fluctuation theory.  We demonstrate that the flat band and van Hove
singularities (VHS) induce profound changes in the superfluid
properties.  In particular, the in-plane effective pair hopping
integral and superfluid density manifest pronounced quantum geometric
effects that significantly bolster superfluidity.  As the chemical
potential crosses the VHS, it displays a nonmonotonic behavior as a
function of interaction strength, reflecting a transition from
hole-like to particle-like pairing.  Most notably, when the Fermi
level resides in the flat band, the diverging DOS drives the BCS
exponential scaling to an \emph{anomalous} power-law dependence of
both the pairing gap and the superfluid transition temperature on
interaction strength, with an exponent $\nu\gtrsim 1$.  This scaling
enhancement underpins a unique power-law temperature dependence of the
superfluid density at low temperatures.  In the strictly 2D limit
($t_z=0$), the pseudogap at $T_\text{c}$ nearly saturates the
zero-temperature gap even at $|g|/t=0.001$, placing the system in a
strong-pseudogap regime that would ordinarily require unitary or
BEC-scale interactions, a direct manifestation of flat-band enhanced
pairing analogous to the strong-correlation physics observed in
magic-angle twisted bilayer graphene~\cite{Cao2018N,Tian2023}.  These findings offer a clear
experimental signature of flat-band superfluidity for future ultracold
atom experiments.

We start with a nearest-neighbor tight-binding model for ultracold Fermi gases in a quasi-two-dimensional Lieb lattice. 
The noninteracting Hamiltonian in momentum space takes the matrix form~\cite{DENG2024AoP} 
\begin{equation*}
  \hat{H}_{\mathbf{k}}=\begin{bmatrix}
    d_k & a_k & b_k\\
    a_k & d_k & 0\\
    b_k & 0 & d_k
  \end{bmatrix},
\end{equation*}
where the off-diagonal elements $a_k=-2t\cos(k_x/2)$ and
$b_k=-2t\cos(k_y/2)$ represent the in-plane hopping, and the diagonal
term $d_k=2t_z(1-\cos k_z)-\mu+2\sqrt{2}t$ describes the out-of-plane
dispersion and the chemical potential $\mu$.  Here, $t$ and $t_z$
denote the in-plane and out-of-plane hopping integrals, respectively,
with the energy zero set at the bottom of the lower band. 
For this study, we set the lattice constant $a=1$ and $t_z/t=0.01$.
Diagonalizing $\hat{H}_{\mathbf{k}}$ yields three bands
$\xi_{\mathbf{k}}^{\alpha}=\alpha\sqrt{2}t\sqrt{2+\cos k_x+\cos
  k_y}+2\sqrt{2}t+2t_z(1-\cos k_z)-\mu$ ($\alpha=0, \pm$), featuring a
flat band ($\alpha=0$) that intersects the dispersive bands
($\alpha=\pm$) at the Dirac points.

To isolate the intrinsic quantum geometric effects, we employ an
effective attractive interaction $U_{\mathbf{k}\mathbf{k}'\alpha\beta}
= g < 0$ that is uniform in the band basis~\cite{DENG2024AoP}.  This
formulation leads to the many-body Hamiltonian
\begin{equation*}
  H=\sum_{\mathbf{k}\alpha \sigma}\xi^{\alpha}_{\mathbf{k}}c_{\mathbf{k}\alpha \sigma}^{\dagger} c^{}_{\mathbf{k}\alpha \sigma}
  + g\!\!\!\sum_{\mathbf{k}\mathbf{k}^{'}\mathbf{q}\alpha\beta}\!\!\! c_{\mathbf{k}_+\alpha\uparrow}^{\dagger} c_{-\mathbf{k}_-\alpha\downarrow}^{\dagger}c^{}_{-\mathbf{k}^{'}_-\beta\downarrow}c^{}_{\mathbf{k}^{'}_+\beta\uparrow}\,,
\end{equation*}
where $\mathbf{k}_\pm \equiv \mathbf{k\pm q}/2$, $c^{}_{\mathbf{k}\alpha\sigma}$ is the annihilation operator for
band $\alpha$ with spin $\sigma=\uparrow,\downarrow$.  By enforcing a
uniform order parameter across all bands~\cite{Chamel2010PRC}, we
isolate the quantum geometric contributions from orbital-selective
pairing effects (see Supplemental Material~\cite{SM} for further
details).

We generalize the pairing fluctuation theory~\cite{Chen1998PRL,Chen2005PR} 
to the multi-band Lieb lattice (see Supplemental Material~\cite{SM}).
The self-energy decomposes into a condensate contribution
$\Sigma_\text{sc}$ and a pseudogap contribution $\Sigma_\text{pg}$
from finite-momentum pairs. The uniform interaction yields a scalar
$T$-matrix $t_\text{pg}$, governed by the pair susceptibility
$\chi(Q)$. The Thouless criterion $1+g\chi(0)=0$~\cite{Thouless1960AoP}
determines $T_\text{c}$, while finite-$Q$ pairs give rise to a pseudogap
$\Delta_\text{pg}$ defined through
$\Delta_\text{pg}^2=-\sum_{Q\neq 0}t_\text{pg}(Q)$.
The total gap satisfies $\Delta^2=\Delta_\text{sc}^2+\Delta_\text{pg}^2$,
with $\Delta_\text{sc}$ vanishing above $T_\text{c}$.
A Taylor expansion of $t_\text{pg}^{-1}(Q)$ yields the effective pair
dispersion and, importantly, allows an unambiguous decomposition of
the in-plane pair hopping $B$ into a conventional contribution
$B_\text{c}$ and a geometric contribution $B_\text{g}$ proportional to
the quantum metric of the dispersive bands~\cite{SM}.
The resulting self-consistent equations read
\begin{equation}
  \label{eq:gap}
  0=\frac{1}{g}+\sum_{\mathbf{k}}\sum_{\alpha=0,\pm}
  \frac{1-2f(E_{\mathbf{k}}^\alpha)}{2E_{\mathbf{k}}^\alpha} \,,
\end{equation}
\begin{equation}
  \label{eq:neq}
  n=\sum_{\mathbf{k}}\sum_{\alpha=0,\pm}\left[ 1-\frac{\xi_{\mathbf{k}}^{\alpha}}{E_{\mathbf{k}}^{\alpha}}+2\frac{\xi_{\mathbf{k}}^\alpha}{E_{\mathbf{k}}^\alpha}f(E_{\mathbf{k}}^\alpha) \right]\,,
\end{equation}
\begin{equation}
  \label{eq:pg}
  \lvert a_0 \rvert \Delta_\text{pg}^2=\sum_{\mathbf{q}}\left(
  1+4\frac{a_1}{a_0}\Omega_\mathbf{q}^0\right)^{-1/2}b(\Omega_{\mathbf{q}})\,,
\end{equation}
where $E_{\mathbf{k}}^{\alpha}=\sqrt{(\xi_{\mathbf{k}}^{\alpha})^2+\Delta^2}$,
$f(x)$ and $b(x)$ are the Fermi and Bose distribution functions, 
$\Omega_\mathbf{q}^0=2B(2-\cos q_x-\cos q_y)+2B_z(1-\cos q_z)$, and
$a_0$, $a_1$ are expansion coefficients~\cite{SM}.
Equations~(\ref{eq:gap})--(\ref{eq:pg}) form a closed set, solved
for $(\mu, \Delta_\text{pg}, T_\text{c})$ with $\Delta_\text{sc}=0$, 
and for $(\mu, \Delta, \Delta_\text{sc}, \Delta_\text{pg})$ at $T<T_\text{c}$.

The superfluid density $(n_s/m)_{\parallel}$ decomposes into
conventional and geometric contributions~\cite{SM}.
The geometric term $(n_s/m)_{\parallel}^\text{geom}$ is
proportional to the quantum metric
$g_{\mu\nu}=\mathrm{Re}\,(\partial_\mu\langle +|)(1-|+\rangle\langle+|)\partial_\nu|+\rangle$,
reflecting interband contributions that arise even with a uniform gap.

\begin{figure}
\centering
\includegraphics[clip,width=3.2in]{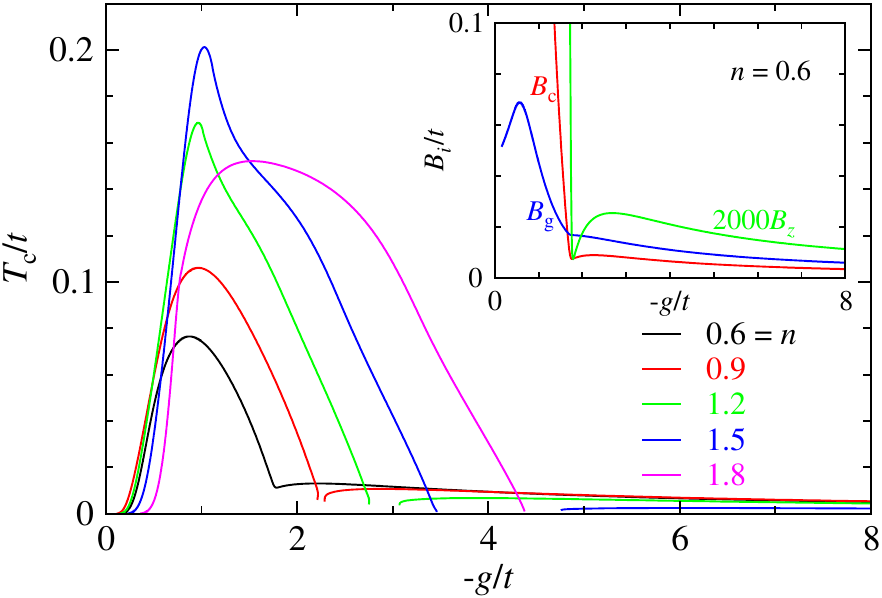}
\caption{ $T_\text{c}$ as a function of the interaction strength
  $-g/t$ for $0.6 \le n \le 1.8$, where the Fermi level lies in the
  dispersive bands.  The inset displays the evolution of the in-plane
  conventional ($B_\text{c}$, red) and geometric ($B_\text{g}$, blue)
  pair hopping integrals, and the magnified out-of-plane component
  ($B_z$, green) for $n = 0.6$.  Note the dominance of $B_\text{g}$
  over $B_\text{c}$ in the strong-coupling BEC regime. }
\label{fig:Tc1}
\end{figure}

Figure~\ref{fig:Tc1} illustrates the behavior of $T_\text{c}$ against
$-g/t$ across the density range $0.6 \le n \le 1.8$.  At low filling
($n \le 1.73$), $T_{\text{c}}$ peaks near unitarity and subsequently
decreases in the BEC regime, consistent with behaviors observed in
three-dimensional (3D) cubic lattices~\cite{Chien2008PRA1}.  Notably,
for $0.65 \le n \le 1.73$, $T_\text{c}$ exhibits a reentrant behavior.
After reaching a maximum, $T_\text{c}$ diminishes and vanishes at
intermediate couplings, before recovering as the interaction
strengthens in the BEC regime.  The suppression of superfluidity
signals the emergence of a pair density wave (PDW) ground state,
driven by strong inter-pair repulsion and large pair sizes at high
densities~\cite{DENG2024AoP}.  Analogous PDW states have been
predicted in 3D lattices~\cite{Chien2008PRA}, two-dimensional (2D)
optical lattices with strong lattice
effects~\cite{Sun2021AdP,Sun2022PRA}, mixed-dimensional
systems~\cite{Zhang2017SR}, and dipolar Fermi gases~\cite{Che2016PRA}.
For $n > 1.73$, superfluidity is entirely absent upon entering the
strong-coupling BEC regime, similar to the 3D lattice
case~\cite{Chien2008PRA1}.

In the BEC regime, $T_\text{c}$ is dominated by the effective pair
hopping integrals.  The inset of Fig.~\ref{fig:Tc1} dissects the
in-plane conventional ($B_\text{c}$) and geometric ($B_\text{g}$)
contributions, as well as the out-of-plane component ($B_z$, magnified
2000$\times$), as a function of $-g/t$ for $n = 0.6$.  In the
weak-coupling BCS regime, the large spatial extent of pairs
facilitates significant overlap, enhancing collective motion and
yielding a large $B_\text{c}$, while $B_\text{g}$ remains negligible.
As inter-pair repulsion intensifies with increasing interaction,
$B_\text{c}$ declines, reaching a minimum at $\mu = 0$, where all
fermions have paired up.  Entering the BEC regime, $B_\text{c}$
initially rises due to the shrinking pair size but subsequently decays
in the strong-coupling regime.  This decay arises from virtual pair
unbinding-rebinding processes that suppress pair
hopping~\cite{Nozieres1985JLTP, Chen1999PRB}.  Crucially, $B_\text{g}$
surpasses $B_\text{c}$ in the deep BEC regime, substantially boosting
$T_\text{c}$.  Finally, $B_z$ follows a trend similar to $B_\text{c}$
but is heavily suppressed by the quasi-2D anisotropy with
$t_z/t=0.01$.

\begin{figure}
\centering\includegraphics[clip,width=3.2in]{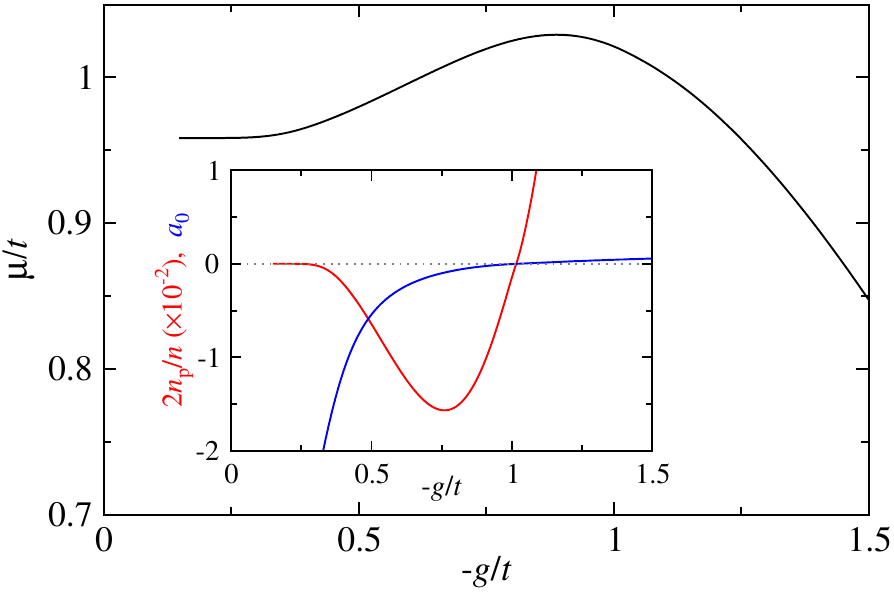}
\caption{ Evolution of the chemical potential $\mu$ as a function of
  $-g/t$ for density $n=1.2$.  The inset displays the corresponding
  pair fraction $2n_\text{p}/n$ and the inverse $T$-matrix coefficient
  $a_0$. }
\label{fig:mu}
\end{figure}

Figure~\ref{fig:mu} presents the behavior of $\mu$, along with the
pair fraction $2n_\text{p}/n$ and the inverse $T$-matrix coefficient
$a_0$ (inset), for $n=1.2$ in the weak-coupling BCS regime, where the
Fermi level resides near the VHS.  Strikingly, $\mu$ exhibits a
\emph{nonmonotonic} dependence on interaction strength, initially
rising with $-g/t$ before decreasing after reaching a maximum.  This
behavior mirrors observations in 2D optical lattices with strong
lattice effects~\cite{Sun2021AdP,Sun2022PRA}.  In the quasi-2D Lieb
lattice, the lower band hosts two VHS's located at
$\varepsilon=(2\sqrt{2}-2)t\approx0.8284t$ and
$\varepsilon=(2\sqrt{2}-2)t+4t_z\approx0.8684t$, defined by the
condition $\nabla\xi_{\textbf{k}}^\alpha=0$~\cite{DENG2024AoP}.  In
the weak-coupling limit, pairing physics is governed by the DOS near
the Fermi surface.  For small $-g/t$, the chemical potential satisfies
$\mu > 0.8684t$, placing the Fermi level above the VHS's where the DOS
exhibits a negative slope.  This configuration induces hole-like
pairing, causing $\mu$ to increase with interaction strength,
analogous to the behavior of a 3D cubic lattice above half-filling.
This phenomenon can be understood analytically through the general
pair density relation
$n_\text{p}=n/2-\sum_{\mathbf{k}\alpha}f(\xi_{\mathbf{k}}^\alpha)$,
where the summation approximates the free fermion density.  As shown
in the inset, $a_0$ is negative at weak interactions, a hallmark of
hole-like pairing~\cite{Sun2021AdP,Sun2022PRA}.  Consequently, the
pair density term $n_{\text{p}} < 0$ leads to an effective increase in
the free fermion density required to conserve total particle number,
thereby pushing $\mu$ above its noninteracting value.  As $-g/t$
increases further, the pairing contribution becomes dominated by the
DOS below the VHS's, causing $\mu$ to decrease.

\begin{figure}
\centering\includegraphics[clip,width=3.3in]{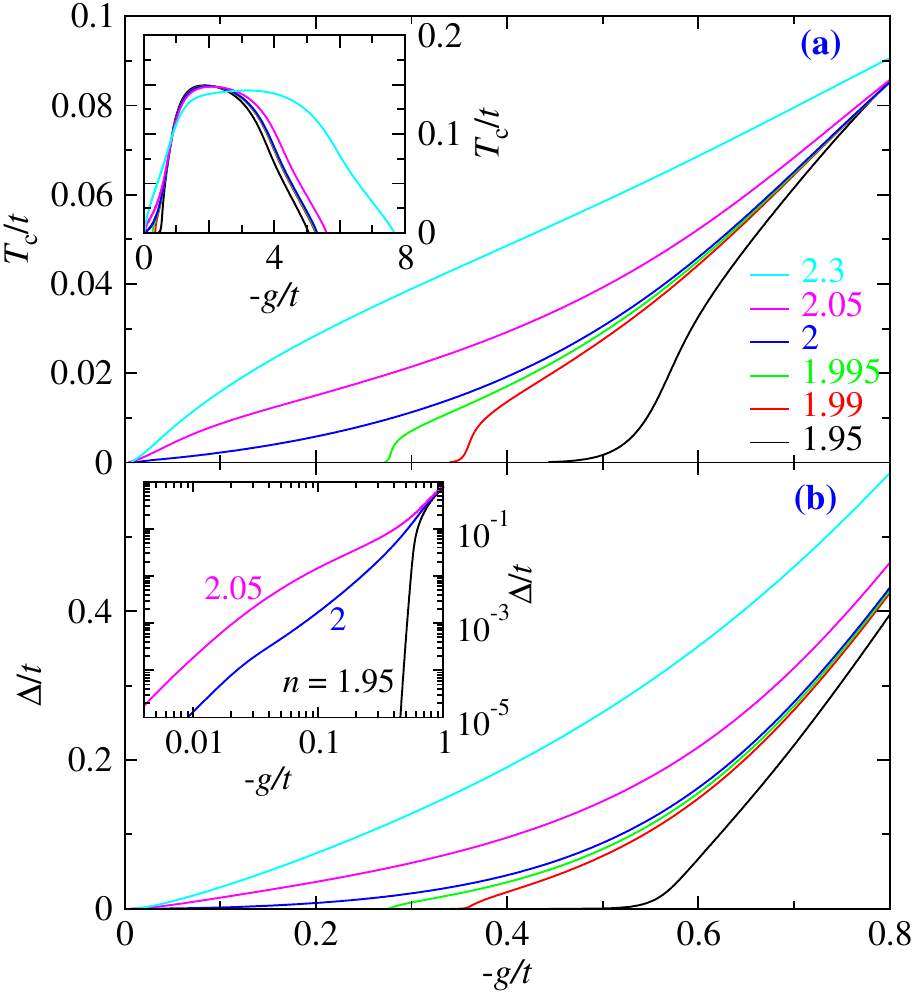}
\caption{ Interaction dependence of (a) $T_\text{c}$ and (b) $\Delta$
  versus $-g/t$ for densities $1.95 \le n \le 2.3$ in the
  weak-coupling regime, where the Fermi level approaches or resides in
  the flat band.  The two insets display (a) the full $T_\text{c}$ curves and
  (b) $\Delta$ on a log-log scale for representative densities $n =
  1.95$ and $2.05$, highlighting the distinct scaling behaviors. }
\label{fig:Tc2}
\end{figure}

Figure~\ref{fig:Tc2} illustrates the dramatic impact of the flat band
on superfluidity when the Fermi level falls in the flat band for $n
\ge 2$.  To visualize the scaling behavior, Panels (a) and (b) present
the evolution of $T_\text{c}$ and $\Delta$ versus $-g/t$ for $1.95 \le
n \le 2.3$. The inset of Fig.~\ref{fig:Tc2}(a) shows the full $T_\text{c}$
curves, while the inset of Fig.~\ref{fig:Tc2}(b) contrasts the
distinct scaling laws on a log-log scale.  For $n < 2$, where the
Fermi level lies below the flat band, $T_\text{c}$ adheres to the
conventional exponential BCS scaling with $-g/t$ at weak interactions.
In stark contrast, for $n > 2$, the quenching of kinetic energy in the
flat band renders the system intrinsically strongly correlated.
Consequently, both $T_\text{c}$ and $\Delta$ undergo a transition from
the exponential BCS dependence to an \emph{anomalous} power-law
behavior.  While the analytical derivation for a strictly flat band
predicts a linear scaling $\Delta \propto |g|$~\cite{DENG2024AoP}, our
numerical results for the quasi-2D system reveal a distinct power-law
scaling $\Delta \propto |g|^\nu$ with an exponent $\nu \gtrsim 1$.
This deviation stems from the finite out-of-plane hopping
$t_z/t=0.01$, which slightly broadens the flat band and weakens the
DOS singularity.  Crucially, this \emph{non-Fermi liquid} scaling
results in significantly enhanced $T_\text{c}$ and $\Delta$ in the
weak-coupling regime compared to the exponentially suppressed BCS
values, signaling a resonant amplification of the superfluid
instability by the flat band.  Analogous transitions from exponential
to power-law scaling have been identified in the magnetic properties
of repulsive Lieb lattices at half filling~\cite{Noda2015PRA}.

\begin{figure}
\centering\includegraphics[clip,width=3.3in]{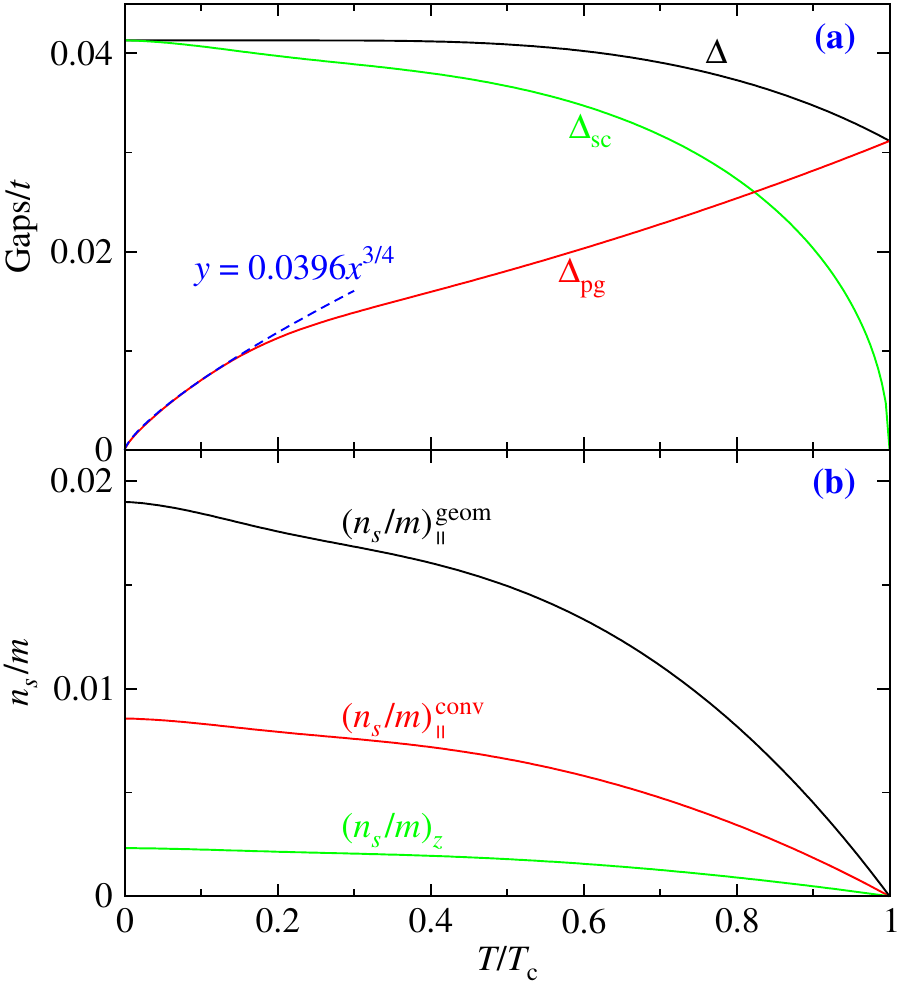}
\caption{ Temperature evolution of (a) the pairing gaps and (b) the
  superfluid density components versus $T/T_\text{c}$ for a flat-band
  filling of $n=2.4$ at weak coupling $-g/t=0.1$.  The blue dashed
  line in (a) indicates a $T^{3/4}$ power-law fit for the pseudogap
  $\Delta_\text{pg}$ at low temperatures.  }
\label{fig:ns}
\end{figure}

Figure~\ref{fig:ns} highlights the thermodynamic signatures of
flat-band superfluidity for $n = 2.4$ at $-g/t=0.1$, where the Fermi
level lies in the flat band.  Panel (a) displays the $T$ dependence of
the total gap $\Delta$, the order parameter $\Delta_\text{sc}$, and
the pseudogap $\Delta_\text{pg}$, together with a $T^{3/4}$ power-law
fit for $\Delta_\text{pg}$ (blue dashed line).  At $T = 0$, the system
is fully condensed with $\Delta_\text{sc} = \Delta$.  As temperature
rises, pairs are excited out of the condensate, so that
$\Delta_\text{sc}$ vanishes at $T_\text{c}$.  While the total gap
$\Delta$ is nearly a constant at low $T$, reflecting exponentially
suppressed quasiparticle excitations, $\Delta_\text{pg}$ exhibits a
remarkable $T^{3/4}$ power law (blue dashed fit), associated with
low-energy bosonic pair fluctuations. The fact $\Delta_\text{pg}$ is
already large at $T_\text{c}$ places the system in the strong-pairing regime, leading to a power-law decrease of $\Delta_\text{sc}$ with increasing $T$, 
typically observed in the unitary or BEC regimes.

The impact of this power-law $T$ dependence extends to the superfluid
transport, as shown in Fig.~\ref{fig:ns}(b).  Through the prefactor
$\Delta_\text{sc}^2$, both the in-plane and out-of-plane superfluid
densities inherit this unconventional thermal behavior; they exhibit a
power-law dependence on $T/T_\text{c}$ at low $T$.  This resembles the
linear $T$ dependence characteristic of nodal $d$-wave
superconductors~\cite{Chen1998PRL,Timusk1999RPP} and should be
contrasted with the exponential behavior of typical fully gapped
$s$-wave superconductors~\cite{Schrieffer2018book}.  Additionally,
while the out-of-plane component $(n_s/m)_z$ is suppressed by the
weak out-of-plane dispersion, the in-plane geometric component
$(n_s/m)^\text{geom}_{\parallel}$ dominates the conventional term
$(n_s/m)^\text{conv}_{\parallel}$, underscoring the capacity of the
flat band's quantum geometry to support robust superfluid transport.
Crucially, this
unusual power-law behavior stems from the strong-pairing nature driven
by the flat band and its quantum geometry rather than from gap nodes,
bearing important physical implications; it cautions against using the
temperature dependence of the superfluid density as a sole indicator
of the order-parameter symmetry in flat-band systems.

\begin{figure}
\centering\includegraphics[clip,width=3.2in]{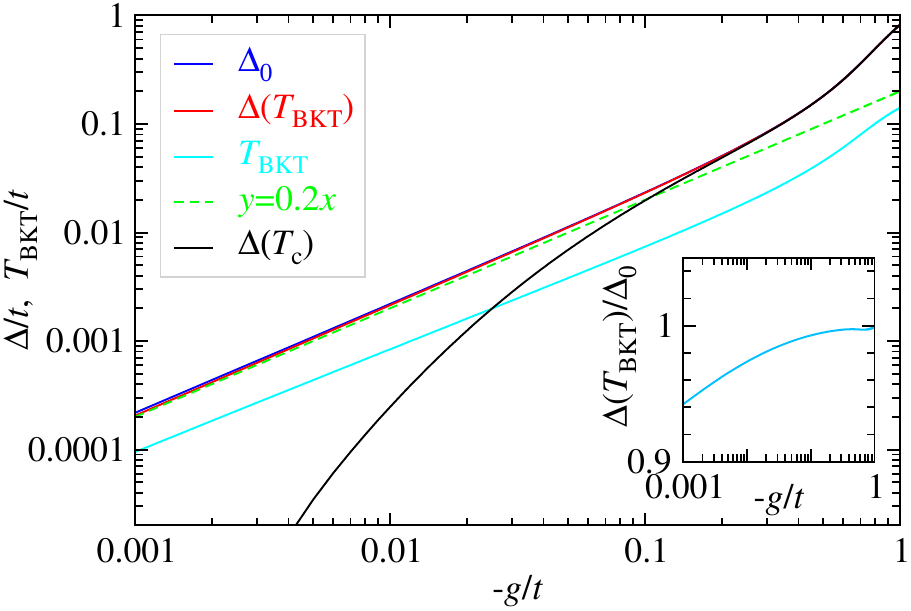}
\caption{ Evolution of the zero $T$ gap $\Delta_0$ (blue) and the
  pseudogap $\Delta(T_\text{BKT})$ (red) as a function of $-g/t$ in
  the 2D limit for $n=2.1$, along with the BKT transition temperature
  $T_\text{BKT}$ (cyan) and the (green) scaling curve $\Delta =
  0.2|g|$. The scaling curve almost overlaps with the two gap curves
  for $-g/t<0.1$. Shown in the inset is the large ratio
  $\Delta(T_\text{BKT})/\Delta_0\lesssim 1$, which reveals a strong-pairing
  flat-band superfluidity in the small $|g|$ limit in 2D.  Also shown
  is $\Delta(T_\text{c})$ for $t_z/t=0.01$.}
\label{fig:2D}
\end{figure}

To isolate the flat-band physics in its purest form, we now turn to
the strictly 2D limit ($t_z=0$), where the DOS singularity is
unbroadened.  In 2D, the superfluid transition is of the
Berezinskii-Kosterlitz-Thouless (BKT) type, governed by vortex-like
phase fluctuations rather than pair breaking. The transition
temperature $T_\text{BKT}$ is determined using the criterion of
critical phase space density ${\cal D}^\text{crit} = 4.9$~\cite{SM}, following
Ref.~\cite{Wang2020}.  Figure~\ref{fig:2D} presents $T_\text{BKT}$,
the zero-$T$ gap $\Delta_0$, and the pseudogap at
$T_\text{BKT}$, $\Delta(T_\text{BKT})$, as functions of $-g/t$ for
$n=2.1$, where the Fermi level lies in the flat band.  For weak
couplings $-g/t < 0.1$, $\Delta_0$, $\Delta(T_\text{BKT})$, and
$T_\text{BKT}$ all follow an approximately linear scaling $\Delta
\propto |g|$ (green curve), consistent with the analytic prediction
for a strictly flat band. Such a linear scaling is a generic feature of flat bands
\cite{PismaZhETF.51.488,Heikkil2011,Kopnin2011PRB,Torma2018,Hofmann2020,Herzog-Arbeitman2022}. We
note, however, this linearity does \emph{not} extend to the strong
coupling regime, where the large gap necessarily brings other
bands into play, and $T_\text{c}$ necessarily decreases toward the BEC regime
after passing a maximum, as shown in Fig.~\ref{fig:Tc2}(a).
Strikingly, the ratio $\Delta(T_\text{BKT})/\Delta_0$ remains close to
unity (inset) throughout the entire coupling range considered, even at
$|g|/t=0.001$.  This places the system deep in a strong-pseudogap
regime, where pairing is essentially saturated and
the transition is driven entirely by the loss of phase coherence---a
hallmark of flat-band superfluidity.  For comparison,
Fig.~\ref{fig:2D} also shows $\Delta(T_\text{c})$ for the quasi-2D case
($t_z/t=0.01$).  Here $\Delta(T_\text{c})$ is strongly suppressed relative to
the 2D result, and on the log-log scale its slope grows monotonically
with decreasing $|g|$, indicating a continuously increasing effective
exponent and hence a crossover toward
conventional exponential BCS behavior in the asymptotic limit
$|g|/t\to 0$, where the effective exponent diverges.

This 2D strong-pseudogap behavior demonstrates that pairing fluctuations
alone can drive a weakly coupled flat-band system into the 
strong-correlation regime, providing a microscopic mechanism for the 
``high-temperature'' superconductivity observed in magic-angle twisted bilayer 
graphene~\cite{Cao2018N,Tian2023}. 

In conclusion, by isolating quantum geometric contributions via a 
band-uniform interaction, we have shown that the in-plane geometric 
pair hopping and superfluid density play a decisive role in enhancing 
$T_\text{c}$ and superfluid transport in the Lieb lattice.  When the Fermi 
level enters the flat band, the BCS exponential scaling collapses into 
anomalous power-law behavior $\Delta, T_\text{c} \propto |g|^\nu$ ($\nu \gtrsim 1$), 
and the superfluid density inherits an unconventional power-law 
temperature dependence, originating from strong pairing rather 
than gap nodes.  In the strictly 2D limit, the pseudogap at $T_\text{c}$ 
nearly saturates $\Delta_0$ even at $|g|/t=0.001$, placing the system 
in a strong-pseudogap regime.  Experimental verification of these 
predictions invites setups beyond the standard on-site interaction 
limit.  Rydberg-dressed Fermi gases offer a promising platform where 
tunable attractive or repulsive non-local interactions can be 
engineered via blue- or red-detuned coupling to approximate the 
effective models discussed 
here~\cite{Balewski_2014,Weckesser2025S}.

This work was supported by the Innovation Program for Quantum Science and Technology (Grant No. 2021ZD0301904).

\bibliography{Ref.bib}

\end{document}


\title{Supplemental Material for ``Quantum Geometry, Anomalous Scaling, and Strong Pseudogap Superfluidity in a Flat-Band Lieb Lattice''}

\maketitle

\section{Pairing Fluctuation Theory for the Multi-Band Lieb Lattice}

\subsection{Uniform interaction and many-body Hamiltonian}

To disentangle the intrinsic quantum geometric contributions from
orbital-dependent interaction effects, we employ an effective
attractive interaction $U_{\mathbf{k}\mathbf{k}'\alpha\beta}=g<0$
that is uniform in the band basis~\cite{DENG2024AoP}.
This leads to the many-body Hamiltonian
\begin{equation*}
  H=\sum_{\mathbf{k}\alpha \sigma}\xi^{\alpha}_{\mathbf{k}}c_{\mathbf{k}\alpha \sigma}^{\dagger} c^{}_{\mathbf{k}\alpha \sigma}
  + g\!\!\!\sum_{\mathbf{k}\mathbf{k}^{'}\mathbf{q}\alpha\beta}\!\!\! c_{\mathbf{k_+}\alpha\uparrow}^{\dagger} c_{-\mathbf{k_-}\alpha\downarrow}^{\dagger}c^{}_{-\mathbf{k^{'}_-}\beta\downarrow}c^{}_{\mathbf{k^{'}_+}\beta\uparrow}\,,
\end{equation*}
where $\mathbf{k}_\pm \equiv \mathbf{k\pm q}/2$. By enforcing a uniform order parameter across all
bands~\cite{Chamel2010PRC}, we preclude orbital-selective pairing
effects, thereby highlighting the quantum geometric contributions.
Furthermore, this formulation streamlines the many-body $T$-matrix
calculation, rendering the analysis of pair susceptibility poles
computationally tractable. 

\subsection{Self-energy decomposition and T-matrix}

We generalize the pairing fluctuation theory, originally developed for
the pseudogap physics in cuprates~\cite{Chen1998PRL} and subsequently
applied to the BCS-BEC crossover in ultracold Fermi
gases~\cite{Chen2005PR}, to finite-temperature Fermi gases in a
multi-band Lieb lattice.  The self-energy
$\Sigma(K)=\Sigma_\text{sc}(K)+\Sigma_\text{pg}(K)$ decomposes into a
Cooper pair condensate term,
$\Sigma_\text{sc}(K)=-\Delta_\text{sc}^2G_0^\text{T}(-K)$, where the
superconducting order parameter $\Delta_\text{sc}$ vanishes above the
superfluid transition temperature $T_\text{c}$ and a pseudogap term
arising from finite-momentum pairs,
$\Sigma_\text{pg}(K)=\sum_{Q\neq0}t_\text{pg}(Q)G_0^\text{T}(Q-K)$,
which persists both above and below $T_\text{c}$.  The effective
uniform interaction yields a scalar pairing $T$-matrix
$t_\text{pg}(Q)=g/[1+g\chi(Q)]$, governed by the pair susceptibility
$\chi(Q)=\sum_{K}\text{Tr}[G(K)G_0^\text{T}(Q-K)]$.  Here, $G(K)$ and
$G_0(K)=(\mathrm{i}\omega-\hat{H}_{\mathbf{k}})^{-1}$ denote the full
and bare Green's functions, respectively.  According to the Thouless
criterion~\cite{Thouless1960AoP}, macroscopic pair occupation at zero
momentum requires $1 + g \chi(0) = 0$, implying that $t_\text{pg}(Q)$
exhibits a sharp peak around $Q=0$.  This allows the approximation
$\Sigma_\text{pg}(K) \approx -\Delta_\text{pg}^2G_0^\text{T}(-K)$,
with the pseudogap defined as
$\Delta_\text{pg}^2=-\sum_{Q\neq0}t_\text{pg}(Q)$.  Consequently, the
total self-energy takes the form
$\Sigma(K)=-\Delta^2G_0^\text{T}(-K)$, where the total pairing gap
satisfies $\Delta^2=\Delta_\text{sc}^2+\Delta_\text{pg}^2$, leading to
the Dyson equation $G^{-1}(K)=G_0^{-1}(K)+\Delta^2 G_0^\text{T}(-K)$.
Throughout, we set $\hbar = k_{\text{B}} = 1$, unity volume, and use
the four-momentum notation $K \equiv (\mathrm{i} \omega_n,
\mathbf{k})$, $Q \equiv (\mathrm{i} \Omega_l, \mathbf{q})$, $\sum_K
\equiv T \sum_n \sum_{\mathbf{k}}$, and $\sum_Q \equiv T \sum_l
\sum_{\mathbf{q}}$, with $\omega_n$ and $\Omega_l$ denoting odd and
even Matsubara frequencies, respectively~\cite{Chen1998PRL}.

\subsection{Self-consistent equations}

 At $T \le T_\text{c}$, the Thouless
criterion~\cite{Thouless1960AoP}, $1+g\chi(0)=0$, yields the gap
equation
\begin{equation}
  \label{eq:gap_sm}
  0=\frac{1}{g}+\sum_{\mathbf{k}}\sum_{\alpha=0,\pm}
  \frac{1-2f(E_{\mathbf{k}}^\alpha)}{2E_{\mathbf{k}}^\alpha} \,,
\end{equation}
where $f(x)$ is the Fermi distribution function, and
$E_{\mathbf{k}}^{\alpha}=\sqrt{(\xi_{\mathbf{k}}^{\alpha})^2+\Delta^2}$
is the quasiparticle dispersion. 

The particle number equation, derived from the constraint $n=2\sum_{K}\text{Tr}G(K)$, reads
\begin{equation}
  \label{eq:neq_sm}
  n=\sum_{\mathbf{k}}\sum_{\alpha=0,\pm}\left[ 1-\frac{\xi_{\mathbf{k}}^{\alpha}}{E_{\mathbf{k}}^{\alpha}}+2\frac{\xi_{\mathbf{k}}^\alpha}{E_{\mathbf{k}}^\alpha}f(E_{\mathbf{k}}^\alpha) \right]\,.
\end{equation}
\subsection{T-matrix expansion and pseudogap equation}

To determine $\Delta_\text{pg}$, we expand the inverse $T$-matrix
$t_\text{pg}^{-1}(Q)$ on the real frequency axis after analytical
continuation ($\mathrm{i}\Omega_l\rightarrow\Omega+\mathrm{i}0^+$).
The expansion takes the form $t_\text{pg}^{-1}(\Omega,\mathbf{q})
\approx a_1\Omega^2+a_0(\Omega-\Omega_\mathbf{q}^0+\mu_\text{p})$,
where $\Omega_\mathbf{q}^0=2B(2-\cos q_x-\cos q_y)+2B_z(1-\cos q_z)$
and the effective pair chemical potential satisfies $\mu_\text{p}=0$
at $T\le T_\text{c}$.  Here, $B$ and $B_z$ represent the in-plane and
out-of-plane effective pair hopping integrals, respectively.  The
coefficients $a_0$, $a_1$, $B$, and $B_z$ are computed from the pair
susceptibility during the Taylor expansion.  Crucially, the uniform
gap function allows an unambiguous decomposition of the in-plane term
$B$ into a conventional contribution $B_\text{c}$ and a geometric
contribution $B_\text{g}$, such that $B = B_\text{c} + B_\text{g}$,
where $B_\text{g}$ is proportional to the quantum metric of the
dispersive bands.  This $T$-matrix expansion yields the pseudogap
equation
\begin{equation}
  \label{eq:pg_sm}
  \lvert a_0 \rvert \Delta_\text{pg}^2=\sum_{\mathbf{q}}\left(
  1+4\frac{a_1}{a_0}\Omega_\mathbf{q}^0\right)^{-1/2}b(\Omega_{\mathbf{q}})\,,
\end{equation}
where $b(x)$ is the Bose distribution function and the pair dispersion
is
$\Omega_\mathbf{q}=(\sqrt{a_0^2+4a_0a_1\Omega_\mathbf{q}^0}-a_0)/2a_1$.
The pair density is approximated by
$n_\text{p}=a_0\Delta^2$~\cite{Sun2021AdP}.

Equations~(\ref{eq:gap_sm})--(\ref{eq:pg_sm}) form a closed set of
self-consistent equations, which we solve for ($\mu$,
$\Delta_\text{pg}$, $T_\text{c}$) with $\Delta_\text{sc}=0$, and for
($\mu$, $\Delta$, $\Delta_\text{pg}$) at $T<T_\text{c}$, where
$\Delta_\text{sc}=\sqrt{\Delta^2-\Delta_\text{pg}^2}$.

\subsection{Superfluid density}

The superfluid density, which quantifies the macroscopic superfluid
transport, provides a definitive probe of the pairing nature through
its low-temperature behavior.  In the Lieb lattice, the in-plane
superfluid density $(n_s/m)_{\parallel}$ decomposes into a
conventional term $(n_s/m)_{\parallel}^\text{conv}$ and a geometric
term $(n_s/m)_{\parallel}^\text{geom}$, satisfying
$(n_s/m)_{\parallel}=(n_s/m)_{\parallel}^\text{conv}+(n_s/m)_{\parallel}^\text{geom}$.
Using the linear response theory adapted for multiband
systems~\cite{Scalapino1992PRL,Kosztin1998PRB,Liang2017PRB,DENG2024AoP},
we derive the expressions for the in-plane conventional and geometric
contributions, along with the out-of-plane component $(n_s/m)_{z}$,
which are given by
\begin{align}
  \label{eq:nsxyc_sm}
  \left(\frac{n_s}{m}\right)_{\parallel}^\text{conv}\!\!=&
  \sum_{\mathbf{k}}\sum_{\alpha=\pm}
  \frac{t^2\Delta_\text{sc}^2}{2{E_{\mathbf{k}}^\alpha}^2}
  \left[\frac{1-2f(E_{\mathbf{k}}^\alpha)}{2E_{\mathbf{k}}^\alpha}
  +f{'}{(E_{\mathbf{k}}^\alpha)}\right]\nonumber\\
    &\times\frac{\sin^2 k_x+\sin^2 k_y}{2+\cos k_x+\cos k_y}\,,\\
  \label{eq:nsxyg_sm}
  \left(\frac{n_s}{m}\right)_{\parallel}^\text{geom}\!\!=&
  \sum_{\mathbf{k}}2\Delta_\text{sc}^2(g_{xx}+g_{yy})\nonumber\\
    \times\!&\left[\left(\!\frac{1-2f(E_{\mathbf{k}}^+)}{2E_{\mathbf{k}}^+}\!-\!\frac{1-2f(E_{\mathbf{k}})}{2E_{\mathbf{k}}}\!\right)\!\frac{\xi_{\mathbf{k}}-\xi_{\mathbf{k}}^+}{\xi_{\mathbf{k}}+\xi_{\mathbf{k}}^+}\right.\nonumber\\
    +\!&\left.\left(\!\frac{1-2f(E_{\mathbf{k}}^-)}{2E_{\mathbf{k}}^-}\!-\!\frac{1-2f(E_{\mathbf{k}})}{2E_{\mathbf{k}}}\!\right)\!\frac{\xi_{\mathbf{k}}-\xi_{\mathbf{k}}^-}{\xi_{\mathbf{k}}+\xi_{\mathbf{k}}^-}\right]\,,\\
  \label{eq:nsz_sm}
  \left(\frac{n_s}{m}\right)_{z}\!\!=&
  \sum_{\mathbf{k}}\!\!\sum_{\alpha=0,\pm}\!\!\!
  \frac{8t_z^2\Delta_\text{sc}^2}{{E_{\mathbf{k}}^\alpha}^2}
  \!\left[\frac{1-2f(E_{\mathbf{k}}^\alpha)}{2E_{\mathbf{k}}^\alpha}
  +f{'}{(E_{\mathbf{k}}^\alpha)}\right]
  \sin^2k_z\,,
\end{align}
where the quantum metric tensor for the upper or lower band is defined
as $g_{\mu \nu}=\text{Re}\, (\partial_\mu\langle {+}|)(1-| + \rangle
\langle +|) \partial_\nu|{+}\rangle$, with $|\pm \rangle$ denoting the
eigenvectors of $\hat{H}_{\mathbf{k}}$ for the upper and lower bands,
respectively.  As evident in Eq.~(\ref{eq:nsxyg_sm}), the geometric term
$(n_s/m)_{\parallel}^\text{geom}$ arises directly from interband
contributions and is proportional to the quantum metric.

\subsection{BKT transition temperature in 2D}

In the strictly 2D limit ($t_z=0$), long-range phase coherence is
destroyed by thermally excited vortex-antivortex pairs.  The
superfluid transition is then of the Berezinskii-Kosterlitz-Thouless
(BKT) type. Here we determine the transition temperature $T_\text{BKT}$ using the criterion of critical phase space density
$\mathcal{D}^\text{crit}=4.9$~\cite{Wang2020}:
\begin{equation}
  \label{eq:BKT}
  \frac{n^{}_{\text{p}}}{M_{\text{B}}}=\frac{{\cal D}_\text{B}^\text{crit}}{2 \pi} T_{\text{BKT}}\,,
\end{equation}
where $n_p$ and $M_\text{B}=1/2B$ are the pair density and effective pair mass,  evaluated at
$T_\text{BKT}$. A more elaborate treatment can be found in Ref.~\cite{Wu2025}.
Note that one could also use the Nelson-Kosterlitz condition to determine  $T_\text{BKT}$. The slight quantitative difference is largely irrelevant for our present purpose.

\bibliography{Ref.bib}